\documentclass[a4paper]{ring}
\usepackage{natbib}
\usepackage{graphicx}
\usepackage[a4paper]{hyperref}
\idline{10006}{1}
\begin{document}
   \title{Exploring X-ray emission and absorption in AGN with XMM-Newton EPIC}

   \author{Ken Pounds}

   \institute{Department of Physics and Astronomy,
              University of Leicester, 
              Leicester LE1 7RH, U.K.}

   \authorrunning{Ken Pounds}  
   \titlerunning{EPIC X-ray Spectra of AGN} 
   \maketitle

\section{Introduction}
The scientific case for XMM \citet{ble84} was accepted by ESA and appears not to have been formally updated since! In
attempting an assessment of the impact of EPIC spectra on our current understanding of AGN it was therefore  necessary to
invent a {\it circa} 1999 science case. With the prominence given to the broad Fe K emission line, based largely on ASCA
results \citet{nan97} the potential study of {\bf strong gravity} would certainly have been a high priority in a
pre-launch XMM science programme. Mapping the {\bf warm absorber and outflows} in general would also have been a priority
target. Given the lack of understanding of the primary X-ray emission process(es) in AGN, I would have argued for a
concerted effort to explore {\bf luminosity and accretion rate} trends. Fourthly, while not having the high energy
capabilities of RXTE or Beppo-SAX, the high sensitivity of EPIC over a broad energy band (~0.2-15 keV) offered the
prospect of exploring spectrally `hard' sources, as in many Type 2 AGN and the dominant AGN component of the CXRB, where
{\bf reflection and/or cold absorption} provide an alternative to an intrinsically flat X-ray continuum.

For the present review, as XMM-Newton approaches the mid-point of (hopefully) a 10-year working life, I have re-examined  
EPIC spectra of 5 bright AGN to assess the impact of data of uniquely high statistical quality on the above science
topics. The data are from observations for which I was a PI (1H0419-577) or Co-I (NGC5548), or have been extracted from
the XMM-Newton archive (NGC4051, PG1211+143, Mkn3). Four of the sources are optical Type 1 Seyfert galaxies, covering a
wide range of X-ray luminosity, while Mkn3 is a bright, nearby Seyfert 2.   

\section{The EPIC spectrum of NGC5548} 
NGC5548 is one of the best-known and most studied Seyfert 1, being a target of every X-ray mission over 30 years. As an
archetypal moderate luminosity AGN, the 125 ks EPIC observation of NGC5548 on 9-12 July 2001  provided an important early
opportunity for a precise measurement of the broad Fe K line indicated in the earlier ASCA observation \citet{nan97}. 
Surprisingly, only a barely resolved, narrow line is seen in the higher quality EPIC data (Fig.1), with an upper limit to
the equivalent width (EW) of a broad line  a factor 5 below the ASCA value. Furthermore, a simultaneous Beppo-SAX
observation allowed the continuum reflection to be quantified, finding a level consistent with the narrow Fe K line EW
$\sim$60 eV. 

Suggested explanations for the unexpected failure to detect reflection from matter close to the black hole in this bright
Seyfert 1 were that the inner accretion disc was highly ionized by the relatively hard incident power law continuum in
NGC5548, or was simply absent at the sub-Eddington accretion ratio of $\sim$5 percent \citet{pou03a}.      
 
\begin{figure}
\centering
\includegraphics[angle=-90,width=7.3cm]{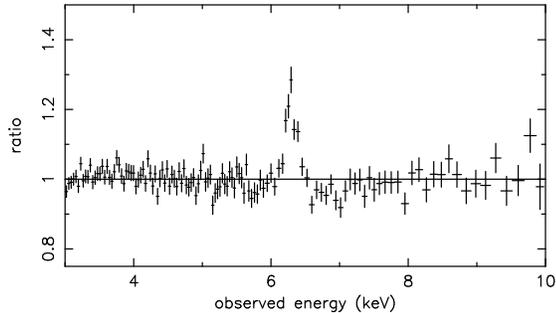}
\caption{Power law fit to the EPIC data of the archetypal Seyfert 1 galaxy NGC5548 showing (only) a narrow Fe K line}
\end{figure}

\section{EPIC observations of NGC4051 in high and low flux states}
Broad-band X-ray continua of AGN are conventionally fitted by a power law, typically of photon index
$\Gamma$$\sim$1.7--2.0, with a `soft excess' evident below $\sim$1--2 keV. The spectral form tends to vary systematically 
with  the flux level of a source, suggesting that spectral variability may help resolve and identify the emission
components. Two EPIC observations of the low luminosity Seyfert 1 galaxy NGC4051 appear particularly promising in this
respect, a long (117 ks) observation on 16/17 May 2001, when the source was bright, being followed by a TOO observation of
52 ks on 22 November 2002, some 20 days into an `extended low' flux interval. These data were recovered from the
XMM-Newton archive and a comparative spectral analysis made \citet{pou04a}. The spectral contrast is remarkable, as
evident in Fig.2 where the integrated bright and faint EPIC (pn) data are plotted against a power law of $\Gamma$=2.

\begin{figure}
\centering
\includegraphics[angle=-90,width=7.3cm]{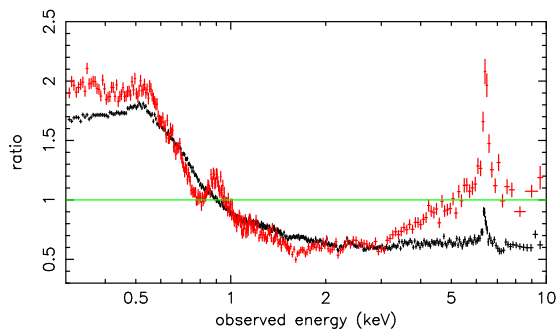}
\caption{EPIC spectra of the low luminosity Seyfert 1 galaxy NGC4051 in the bright (black) and faint flux (red) states plotted against a power law 
of $\Gamma$=2}
\end{figure}

While the bright state spectrum is well fitted at 2--10 keV by the assumed $\Gamma$=2 power law (again with only a narrow
Fe K line), with a smooth soft excess at lower energies, the low flux spectrum is dramatically different. The much harder
continuum $\ga$3 keV and more pronounced (but $\sim$constant flux) narrow Fe K line can be explained by  retaining the
strength of the reflected spectrum while the direct continuum flux is $\sim$5 times fainter in the second observation,
implying the reflecting matter is $\ga$20 light days from the continuum source (outer disc or torus?). Interestingly, a
2--10 keV power law fit excluding the 5--7 keV band (the convention used in most ASCA studies) suggests the presence of a
strong and broad Fe K line in the low flux state spectrum. While this cannot be excluded  with the present data, it may
seem counter-intuitive for a feature requiring reflection from the innermost accretion disc to be evident only when the
intrinsic continuum source is almost switched off.

An alternative suggested by enhanced structure throughout the low state EPIC spectrum is that these features are
predominently due to the imprint of absorption on the underlying continuum. Fig.3 reproduces such a model fit to the low
flux EPIC spectrum of NGC4051, with the intrinsic power law partially covered by a warm absorber, modelled  by XSTAR
\citet{kal96}, of ionization parameter $\xi$$\sim$20 and  column densities N$_{H}$$\sim$$10^{22}$cm$^{-2}$ and
N$_{H}$$\sim$$10^{23}$cm$^{-2}$. It is notable that the intrinsic power law in this low flux state fit, though much
fainter, has the same spectral index $\Gamma$$\sim$2 as in the high flux state.
    
\begin{figure}
\centering
\includegraphics[angle=-90,width=7.3cm]{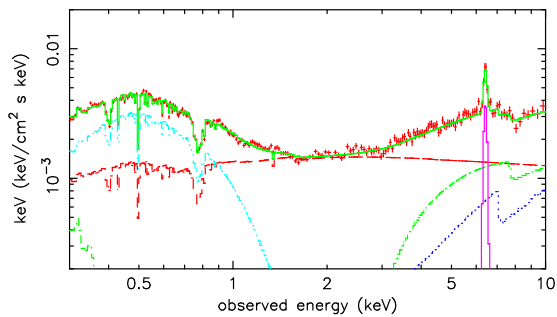}
\caption{Model fit to the low state EPIC spectrum of NGC4051 with the power law continuum
of $\Gamma$=2 partially absorbed by ionized gas of column densities N$_{H}$$\sim$10$^{22}$cm$^{-2}$ (red)
and N$_{H}$$\sim$$10^{23}$cm$^{-2}$ (green). Other components in the fit are a narrow Fe K emission line (magenta),
continuum reflection (dark blue) and blackbody (light blue)}
\end{figure}

\section{The highly variable Seyfert galaxy 1H0419-577}
A series of XMM-Newton observations of the luminous Seyfert 1 galaxy 1H0419-577 was carried out over the period September
2002 to September 2003, this source being chosen for detailed study as one of the most spectrally variable AGN.  Five
observations, each of nominally 15 ks, were made at 3-monthly intervals, during which time the source cooperated fully in
exhibiting a wide range of flux states (Fig.4). As with NGC4051 the lower flux state EPIC spectra show more structure. An 
analysis of these data \citet{pou04c}
explored the spectral variability in terms of the {\it difference spectra}, obtained by subtracting the extreme low state
EPIC data from each higher flux spectrum.  By this means it was established that the main spectral change was due to the
varying strength of the  power law component, with the rather steep index of $\Gamma$$\sim$2.4 emphasising the spectral
softening with increasing flux evident in Fig.4. 

A second important result from this EPIC study of 1H0419-577 was that a substantial column density of cold gas, imprinting
absorption on the spectra in the lower flux states, progressively `thinned out' as the source brightened (Fig.5).
Modelling the absorption with XSTAR, on the same $\Gamma$$\sim$2.4 power law continuum, found a column density of
N$_{H}$$\sim$$10^{22}$cm$^{-2}$ in the mid-low state difference spectral fit, falling to N$_{H}$$\sim$$2\times 
10^{21}$cm$^{-2}$ in the intermediate state fit, and to a negligible value in the extreme high state spectrum \citet{pou04b}.

\begin{figure}
\centering
\includegraphics[angle=-90,width=7.3cm]{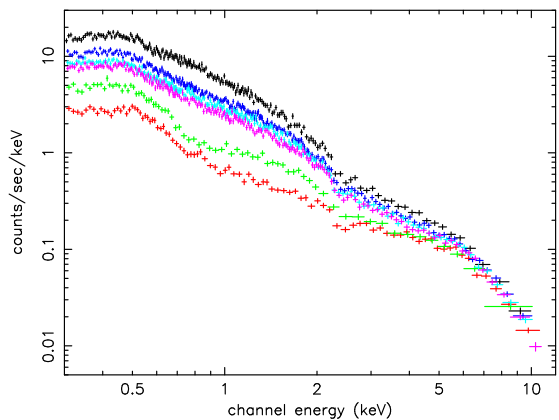}
\caption{EPIC pn spectral data of 1H0419-577 from the observations of December 2000 (high state:black), September 2002 
(low state:red), December 2002 (mid-low state:green),
March 2003 (mid-high state:purple),
June 2003 and September 2003 (intermediate state:light blue and magenta)}
\end{figure}

Underpinning this {\it difference spectrum} analysis is the implication that there is a quasi-constant component or
components dominating the lowest flux state. A conventional fit to the low flux spectrum of 1H0419-577 reveals a hard power 
law, with an
extreme Fe K line-like feature, and a strong soft excess. However, just as in modelling of the low state EPIC data for
NGC4051, continuum absorption by a large column of low ionization matter offers an equally good fit.

\begin{figure}
\centering
\includegraphics[angle=-90,width=7.3cm]{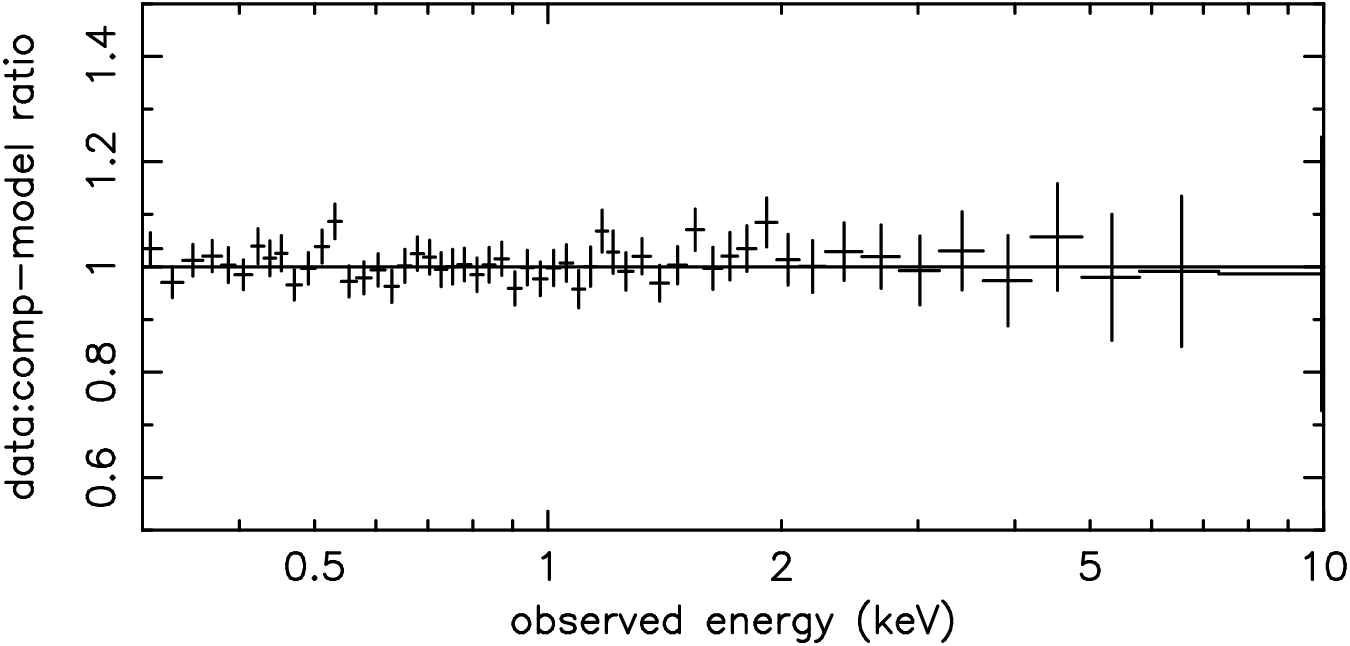}
\centering
\includegraphics[angle=-90,width=7.3cm]{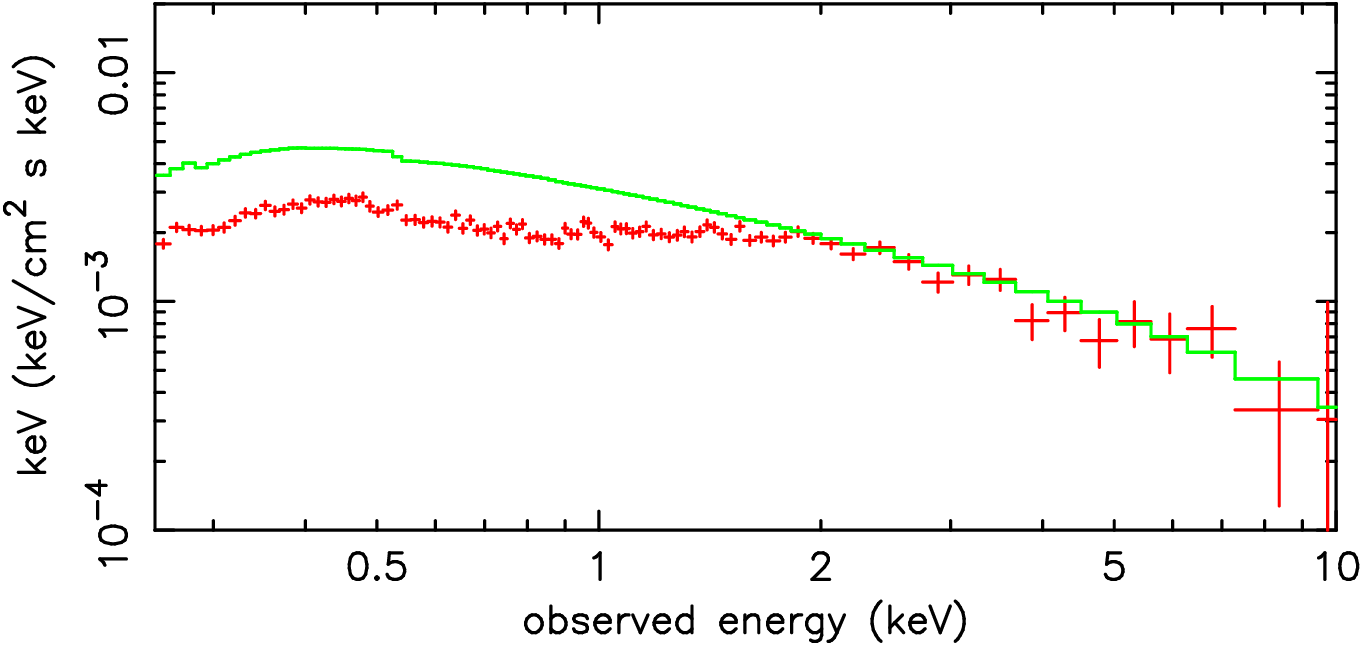}
\centering
\includegraphics[angle=-90,width=7.3cm]{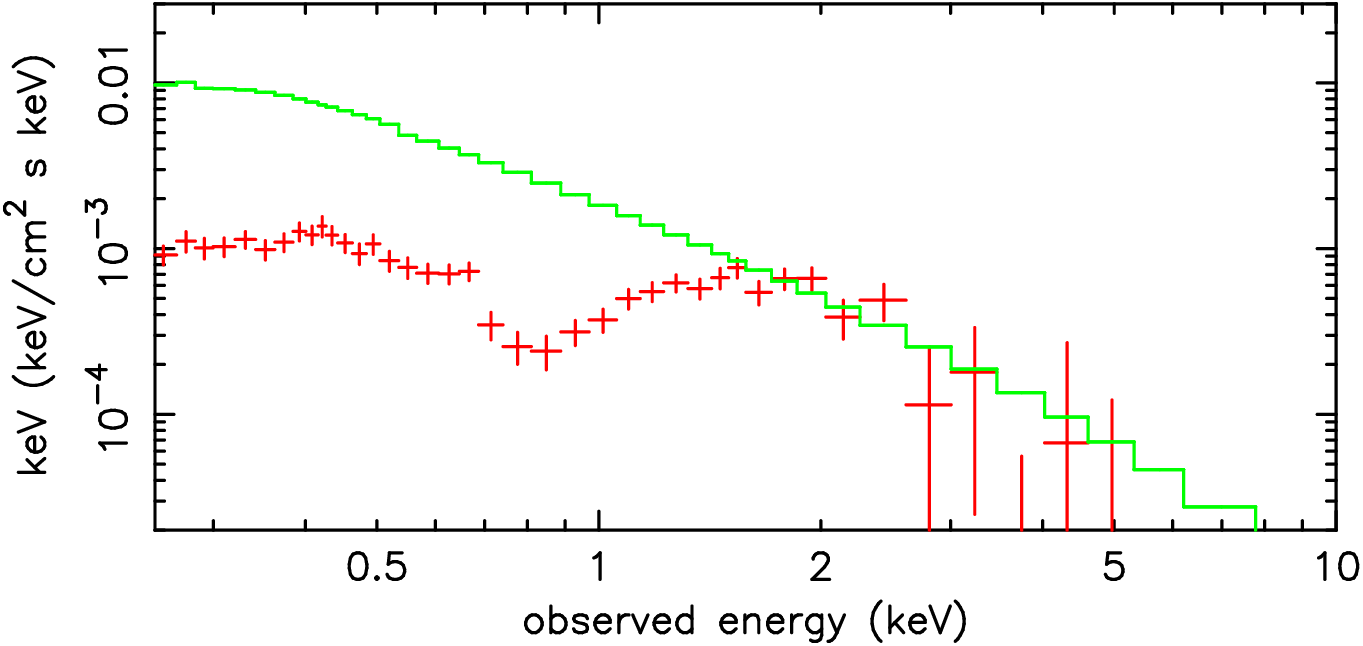}
\caption{(top) High state difference spectral data of 1H0419-577 plotted against a power law of $\Gamma$$\sim$2.4;
(middle) intermediate state difference spectral data of 1H0419-577 (red) compared with the cut-off power law
model with the cold absorption removed (green); (lower) mid-low state difference spectral data of 1H0419-577 (red) 
compared with the cut-off power law
model with the cold absorption removed (green). See text for details}
\end{figure}

A third important outcome of the above EPIC study of 1H0419-577 was that the `soft excess', for so long a feature  of AGN
X-ray spectra, is seen to be - in substantial part - an artifact of absorption on the underlying continuum. This finding
supports a recent proposal \citet{gie04} to address the long-standing difficulty of physically matching the  soft excesses
in AGN with thermal emission from the accretion disc. We now see, at least for 1H0419-577, the soft excess is  a
combination of a steeper (than the canonical $\Gamma$$\sim$1.8-2) power law, attenuated below $\sim$1 keV by strong and
variable absorption, and a quasi-constant  soft component. It seems quite likely that this {\it true} soft X-ray emission
is associated with the ionized outflow more generally observed in absorption in Type 1 AGN.

\section{High velocity outflows - a major XMM-Newton discovery}
The improved resolution of ASCA spectra confirmed the presence of `warm absorbers' in many AGN, with spectral fits
suggesting typical column densities of N$_{H}$$\sim$$10^{21}-10^{22}$cm$^{-2}$ and ionization parameter $\xi$$\sim$10-50
imposing strong absorption of OVII, OVIII and Fe-L ions on the soft X-ray continuum. High resolution RGS and Chandra
spectra have confirmed and quantified such `warm outflows' in many AGN. The greater sensitivity of EPIC at higher energies
has provided much improved spectral data in the Fe K band $\ga$7 keV, showing that significant absorption is common there
too in AGN spectra. This is particularly important in providing a measure of absorption by more highly ionized gas,
transparent at lower energies. The most remarkable discoveries to date have been of extremely deep Fe K absorption edges
in 2 Narrow Line Seyferts 1H0707-495 \citet{gal04} and IRAS13224-3809 \citet{bol03} and of high energy absorption lines
indicating high velocity outflows. 

Most of the early claims for high velocity outflows in AGN X-ray spectra have been based on XMM-Newton data, although 
recently there have been at least 2 similar reports from Chandra. [A current listing of 10 AGN showing X-ray evidence for
high velocity outflows is included in the MPE Website covering the Ringberg meeting, where a common factor appears to be
AGN with a high accretion ratio.] The luminous Seyfert 1 PG1211+143 remains one of the best examples of this phenomenon, as
evidence for an ionized outflow at v$\sim$0.08--0.1c is seen in both EPIC and RGS data \citet{pou03b},\citet{pou05a}.
Fig.6 reproduces the EPIC data where narrow  absorption features are identified with Lyman $\alpha$ resonance lines of
FeXXVI, SXVI and MgXII. 

The wider significance of these high velocity winds, in comparison with the gentler `breeze' more often seen in low
energy absorption spectra, is underlined by noting that the mechanical energy involved can be comparable to the accretion
energy  \citet{pou03b}.  Thus, integrated over $\sim$$10^{8}$ years, such a wind could have a major impact on the growth
of the host galaxy \citet{kin03}.

\begin{figure}
\centering
\includegraphics[angle=-90,width=7.3cm]{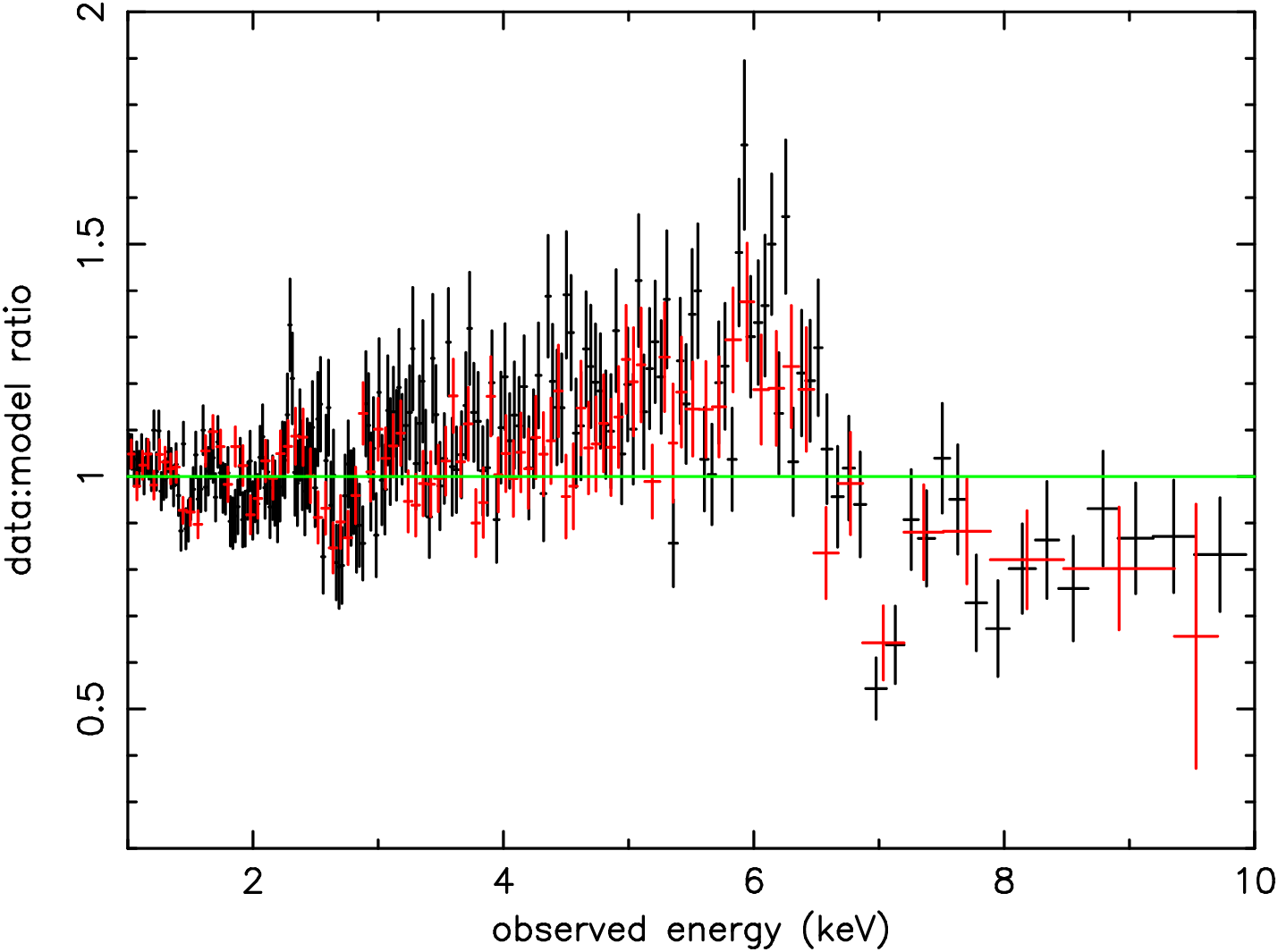}
\centering
\includegraphics[angle=-90,width=7.3cm]{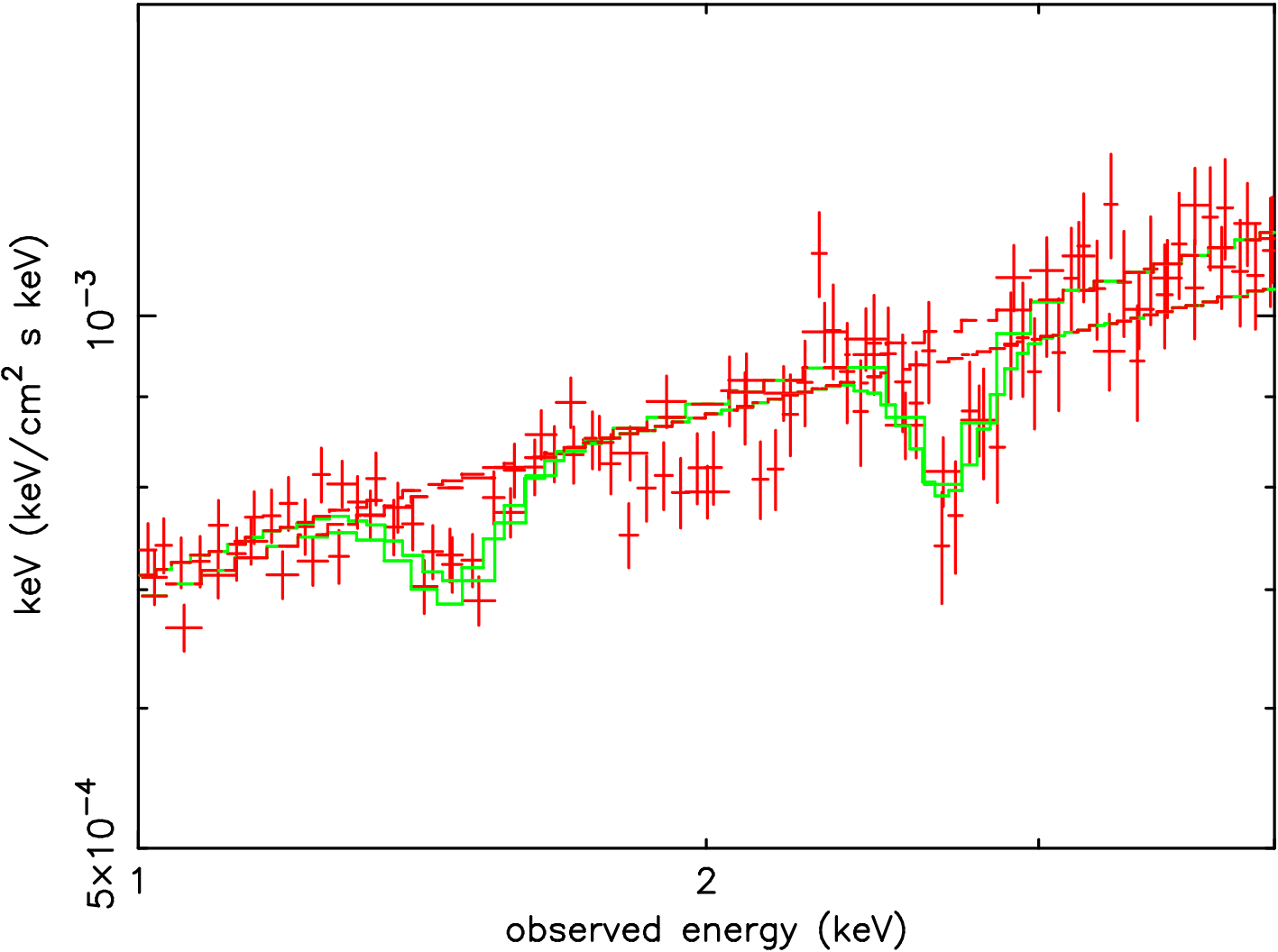}
\caption{Absorption lines in the EPIC spectrum of PG1211+143 identified (top) with highly ionized Fe `blue-shifted' 
by v$\sim$0.1c; (lower) with highly ionized S and Mg also
`blue-shifted' by v$\sim$0.1c}
\end{figure}
  
\section{An EPIC observation of the nearby Seyfert 2 galaxy Mkn3}

Mkn3 is a low redshift (z=0.0135) Seyfert 2 for which ASCA \citet{tur97} and Beppo-SAX \citet{cap99} observations had shown a
hard, reflection-dominated spectrum. A 105 ks XMM-Newton observation on 19/20 October 2000 was retrieved from the archive.
The EPIC spectral data (Fig.7) extends to $\sim$15 keV indicating a very hard spectrum; in addition a strong Fe K line and
several other prominent spectral features are seen. The statistical quality and extended bandwidth of the EPIC data are well
fitted with the reflection-dominated model proposed for the ASCA and Beppo-SAX spectra, though now the model parameters are
better constrained. The intrinsic continuum, with a typical Seyfert power law index $\Gamma$$\sim$1.8, is seen through a
near-Compton-thick absorber (N$_{H}$$\sim$$1.3\times10^{24}$cm$^{-2}$), with the hard spectrum at $\sim$3-8 keV being
dominated by continuum  reflection and fluorescent line emission. 

In addition to the strong K$\alpha$ and $\beta$ lines of Fe, fluorescent emission from the same re-processing matter is
detected - at strengths consistent with solar abundances - for Ni, Ca, Ar, S, Si and Mg. A soft excess extending above a low
energy extrapolation of the 3-15 keV model fit (Fig.8) is then impressively well resolved by the EPIC detectors (eg Fig.9).
The clarity of this plot underlines the spectroscopic potential of the MOS camera, in particular, having a resolution of 
$\sigma$$\sim$34 eV at 1.5
keV.

\begin{figure}
\centering
\includegraphics[angle=-90,width=7.3cm]{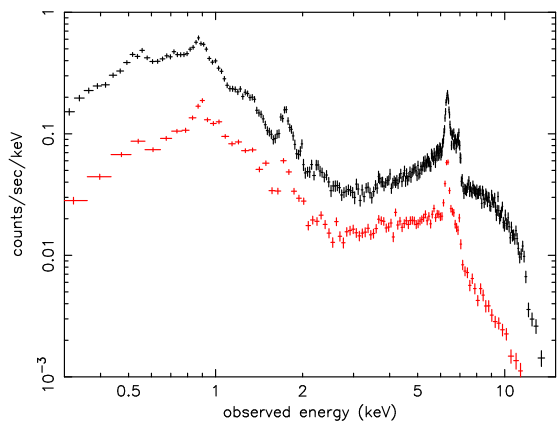}
\caption{Background-subtracted spectra from the pn (black) and MOS (red) observations of Mkn3}
\end{figure}

The particularly strong Fe K$\alpha$ line is intriguing, having an energy of 6.43$\pm$0.01 keV in both pn and MOS cameras.
In comparison to a laboratory energy (weighted mean of the K$\alpha$ doublet) of 6.400 keV, the data suggest a blueshift -
in velocity terms - of 1400$\pm$450 km s$^{-1}$. The sharp Fe K absorption edge, presumably from the same matter, lies
close to 7.1 keV excluding ionization as the origin of the line energy shift. This result fits uncomfortably with an
origin of the Fe K line from the unobscured far, inner wall of a torus, unless located on much less than the conventional
$\sim$pc scale \citet{pou05}. 

Resolving the soft emission from Mkn3 with EPIC (eg Fig.9) also reveals strong resonance emission lines of highly ionized
N, O, Ne, Mg, Si and S. Higher resolution spectra from the RGS confirm these identifications and find the soft spectrum to
be consistent with a photo-ionized/photo-excited outflow, as previously reported from a Chandra study \citet{sak00}, at a
projected velocity of  $\sim$800 km s $^{-1}$. The integrated luminosity in this soft X-ray component is $\sim$1 percent
of the absorption-corrected power law continuum and represents only that part of the outflow extending above the  Compton
thick screen. The intriguing question as to how much greater would the soft X-ray emission be were  Mkn 3 viewed at a
smaller inclination (ie as a Seyfert 1) is interesting in relation to the origin of the quasi-constant soft X-ray emission
in cases such as the low flux states of NGC4051 and 1H0419-577.  

\begin{figure}
\centering
\includegraphics[angle=-90,width=7.3cm]{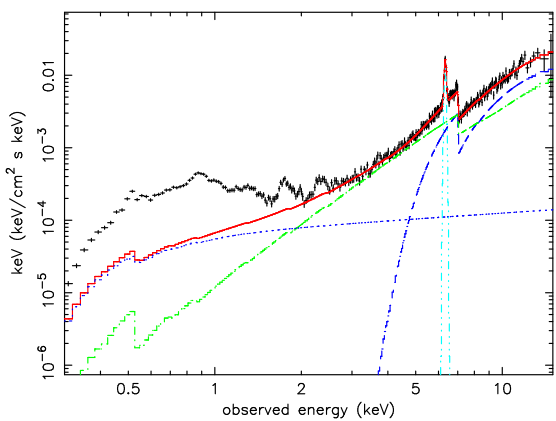}
\caption{Unfolded model fit to the EPIC spectrum of Mkn3 showing the absorbed power law (dark blue) and
reflection (green) components to the hard X-ray continuum, with a strong Fe K emission line (light blue).
Extrapolating this fit below 3 keV (red line) reveals a soft excess, only a small part of which can be attributed to an
unabsorbed fraction of the intrinsic power law (blue dots)}
\end{figure}

\begin{figure}
\centering
\includegraphics[angle=-90,width=7.3cm]{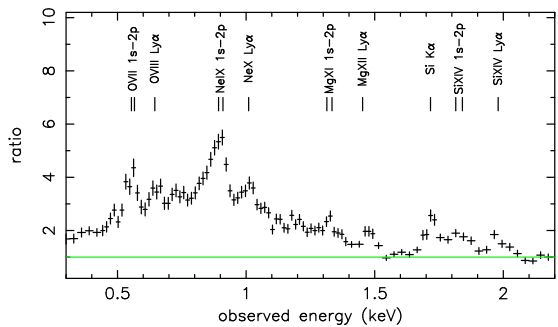}
\caption{Resolving the soft excess in the MOS spectrum in terms of principal emission lines of highly ionized
O, Ne, Mg and Si, together with fluorescent SiI K$\alpha$}
\end{figure}

\section{Summary and Conclusions}
EPIC is proving to be a powerful instrument for the study of AGN X-ray spectra. Its sensitivity over a uniquely broad
energy band has provided the best data yet for deconvolving the primary and secondary emission components, with the
ability to follow flux-linked changes being particularly instructive. While not having the high spectral resolution of
grating instruments, the CCD resolving power on EPIC is impressive - and probably could be more fully exploited (a
potential that improved confidence in instrument calibration should help realise). Mkn3 is a good demonstration of
the spectroscopic capabilities of EPIC, where the high energy response allows the obscured power law and strong reflection
components to be unambiguously resolved, while providing a clear resolution of the line-dominated soft X-ray
emission.

The complementarity of EPIC and the RGS on XMM-Newton is well illustrated in the study of ionized outflows in AGN. While
the high spectral resolution of the RGS is essential to determine the line widths and velocities of a few 100 km s$^{-1}$
seen in absorption in many AGN, EPIC data has demonstrated that much more highly ionized gas - transparent in the soft
X-ray band -  is often present. In addition, the combination of RGS and EPIC spectra is important in demonstrating the high
velocity and energetically important outflows in objects such as PG1211+143 and PDS456 \citet{ree03}.
A further, less obvious, complementary strength of EPIC arises in quantifying absorption in cold matter, or where the
absorbing gas is highly turbulent, neither case yielding the narrow features well suited to the RGS. 

In conclusion, what has been learned on the priority science topics listed in our pre-launch science case update?
Unfortunately, it seems the {\bf broad Fe K line} remains controversial as a probe of strong gravity. As in the EPIC observation of
NGC5548, only a narrow Fe K emission line, with a strength and profile consistent with an origin in cold matter distant from
central Black Hole, is seen in a majority of EPIC spectra of AGN. While a recent survey of XMM observations (Reeves
and Nandra, unpublished) showed a diversity of Fe K line profiles, with several having excess emission to the `red' or `blue'
sides of the 6.4 keV line, the evidence for a relativistically broadened line containing
information on the effects of strong gravity and black hole spin remains unclear. The tendency, as in NGC4051 and 1H0419-577,
for a broad `line-like' feature to be most evident in low flux states supports an alternative description, where the observed
feature is an imprint of enhanced absorption by cold or weakly ionised matter on the power law continuum. The fact that this
ambiguity remains despite the careful analysis of several high quality EPIC spectra, eg MCG-6-30-15 \citet{fab02}
\citet{vau04}, underlines the key importance of future observations being able to fully map the effects of such
absorption. Meanwhile we can only speculate that the Fe K signature of re-processing in matter that {\it must} exist
close to the Black Hole in any radiatively efficient AGN is often diluted by ionization or is obscured by other accreting or
outflowing matter.     

While the location of such absorbing matter is uncertain, perhaps residing in the outer layers of the accretion
disc or base of a jet, XMM-Newton spectra have revealed strong {\bf ionized outflows} in many AGN, both in absorption and
emission. Perhaps the  most significant new discoveries, primarily led by EPIC data, relate to the evidence for highly ionized
winds moving at $\sim$0.1--0.2c. Simply to detect the tell-tale absorption lines of FeXXV or XXVI requires column densities of
order N$_{H}$$\sim$$10^{23}$cm$^{-2}$, which translate into mass outflow rates comparable to, or exceeding, the accretion
rates for the respective AGN unless the flow cone is narrow. A high velocity then makes these flows also important
energetically, offering a natural link between black hole and host galaxy growth as implicated in observational correlation
between the black hole mass and the velocity dispersion in the galactic bulge \citet{fer00}, \citet{tre02}. The fact that the
growing list of AGN with reported {\bf high velocity outflows} may mostly (if not all) be {\bf accreting at the Eddington
rate} is consistent with matter being ejected - at the local escape velocity - from the inner disc region \citet{Kin03}.

The repeated XMM-Newton observations of another high luminosity Seyfert 1, 1H0419-577, provides a unique data base to study
large-scale {\bf spectral variability}. The analysis outlined here finds the dominant variable to be a relatively steep power
law component, as was previously found for MCG-6-30-15 \citet{fab03}, suggesting this may be a common factor, if not always
so obvious. Modelling difference spectra strongly suggests - again - the presence of substantial {\bf cold absorbing matter},
the column density (or covering factor) of which falls as the source flux increases.

Finally, the XMM-Newton observation of Mkn3, sitting in the data archive for over 4 years, is an impressive demonstration  of
the  diagnostic power of EPIC. Not only do those data allow the direct observation of a `normal' Seyfert 1 power law, seen
through a near-Compton-thick absorber, the {\bf strong reflected continuum} and associated fluorescent line spectrum are well
resolved. In addition the strong Fe K$\alpha$ line profile is measured with sufficient precision to provide an interesting
indication that much of  the re-processing matter lies at a smaller radius than the expected inner boundary of a molecular
torus. Here again, improved confidence in the in-flight calibration data for EPIC would allow the strong Fe K line to be an
important probe of the circumnuclear matter in Type 2 AGN.

\begin{acknowledgements}
It is a pleasure to pay tribute to the outstanding efforts of all those colleagues who designed and built EPIC
and those who continue to maintain it as a unique instrument for X-ray Astronomy.
My own particular thanks go to the Leverhulme Trust 
for the support of an Emeritus Research Fellowship.      
\end{acknowledgements}

\bibliographystyle{aa}

\end{document}